\documentclass[aps,prd,superscriptaddress,preprintnumbers]{revtex4}
\usepackage{graphicx}

\newcommand*{\ie}{\textit{i.e.},\ }

\newcommand*{\gevc}{\,\textrm{GeV/c}}

\newcommand*{\tev}{\,\textrm{TeV}}
\newcommand*{\pt}{$p_T$}
\newcommand*{\bigpt}{$P_T$}
\newcommand*{\UE}{``underlying event"}
\newcommand*{\BBR}{``beam-beam remnants"}

\newcommand*{\TR}{``transverse"}
\newcommand*{\tmax}{``transMAX"}
\newcommand*{\tmin}{``transMIN"}
\newcommand*{\tdif}{``transDIF"}
\newcommand*{\LJ}{``leading jet"}
\newcommand*{\BB}{``back-to-back"}

\newcommand*{\ptone}{$P_T({\rm jet}\#1)$}

\newcommand*{\Jone}{jet\#$1$}
\newcommand*{\Jtwo}{jet\#$2$}
\newcommand*{\Jthree}{jet\#$3$}

\newcommand*{\ptsum}{$PT{\rm sum}$}
\newcommand*{\etsum}{$ET{\rm sum}$}
\newcommand*{\ptcut}{$p_T\!>\!0.5\,{\rm GeV/c}$}
\newcommand*{\etcut}{$E_T\!>\!0.1\,{\rm GeV}$}
\newcommand*{\etacut}{$|\eta|\!<\!1$}

\newcommand*{\avept}{$\langle\!p_T\!\rangle$}
\newcommand*{\etaphi}{$\eta$-$\phi$}
\newcommand*{\delphi}{$\Delta\phi$}

\newcommand*{\delphicut}{$|\Delta\phi|>150^\circ$}
\newcommand*{\nden}{$dN_{chg}/{d\phi}{d\eta}$}
\newcommand*{\ptden}{$dPT_{sum}/{d\phi}{d\eta}$}
\newcommand*{\etden}{$dET_{sum}/{d\phi}{d\eta}$}
\begin{document}
\preprint{CDF/ANAL/CDF/PUBLIC/7822}
\title{PYTHIA Tune A, HERWIG, and JIMMY \\ in Run 2 at CDF}
\author{Rick Field}
\affiliation{
Department of Physics, University of Florida,
Gainesville, Florida, 32611, USA \\ \rm{(for the CDF Collaboration)}}
\author{R. Craig Group}
\thanks{To appear in the proceedings of the HERA-LHC workshops.}
\affiliation{
Department of Physics, University of Florida, 
Gainesville, Florida, 32611, USA \\ \rm{(for the CDF Collaboration)}}
\date{September 1, 2005}

\begin{abstract}
We study the behavior of the charged particle (\ptcut, \etacut) and energy (\etacut) components of 
the \UE\ in hard scattering proton-antiproton collisions at $1.96\tev$. The goal is to produce data on 
the \UE\ that is corrected to the particle level so that it can be used to tune the QCD Monte-Carlo 
models without requiring CDF detector simulation. Unlike the previous CDF Run $2$ \UE\ analysis which 
used JetClu to define ``jets" and compared uncorrected data with the QCD Monte-Carlo models after 
detector simulation (\ie CDFSIM), this analysis uses the MidPoint jet algorithm and corrects the 
observables to the particle level.  The corrected observables are then compared with the QCD Monde-Carlo 
models at the particle level (\ie generator level).   The QCD Monte-Carlo models include PYTHIA Tune A, 
HERWIG, and a tuned version of JIMMY.
\end{abstract}

\maketitle

One can use the topological structure of hadron-hadron collisions to study the \UE\ 
\cite{CDF:Field,Field:DPF2000,Huston:DPF2000}.  The direction of 
the leading calorimeter jet is used to isolate regions of \etaphi\ space that are sensitive to the \UE. As 
illustrated in Fig.~1, the direction of the leading jet, \Jone, is used to define correlations in the azimuthal 
angle, \delphi.  The angle $\Delta\phi=\phi-\phi_{\rm jet\# 1}$ is the relative azimuthal angle between a 
charged particle (or a calorimeter tower) and the direction of \Jone.  The \TR\ region is perpendicular 
to the plane of the hard $2$-to-$2$ scattering and is therefore very sensitive to the \UE. We restrict ourselves 
to charged particles in the range \ptcut\ and \etacut\ and calorimeter towers with \etcut\ and \etacut,  but allow 
the leading jet that is used to define the \TR\ region to have $|\eta(jet\# 1)|<2$.   Furthermore, we 
consider two classes of events.  We refer to events in which there are no restrictions placed on the second 
and third highest \bigpt\ jets (\Jtwo\ and \Jthree) as \LJ\ events.  Events with at least two jets 
with $P_T>15\gevc$ where the leading two jets are nearly \BB\ (\delphicut) with $P_T(jet\#2)/P_T(jet\#1)>0.8$ 
and $P_T(jet\#3)<15\gevc$ are referred to as \BB\ events. ``Back-to-back" events are a subset of the \LJ\ events.  The 
idea is to suppress hard initial and final-state radiation thus increasing the sensitivity of 
the \TR\ region to the  \BBR\ and the multiple parton scattering component of the \UE. 

\begin{figure}[htbp]
\begin{center}
\includegraphics[scale=0.6]{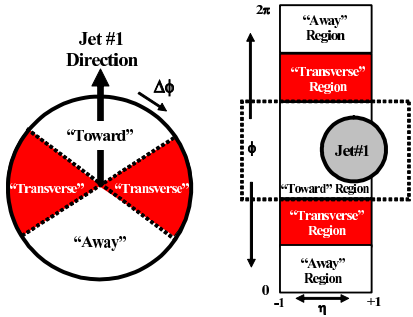}
\caption{
Illustration of correlations in azimuthal angle $\phi$ relative to the direction of the 
leading jet (MidPoint, $R=0.7$, $f_{merge}=0.75$) in the event, jet\#$1$.  The angle 
$\Delta\phi=\phi-\phi_{\rm jet1}$ is the relative azimuthal angle between charged 
particles and the direction of \Jone.  The \TR\ region is defined 
by  $60^\circ<|\Delta\phi|< 120^\circ$ and \etacut.  We examine charged particles in the 
range \ptcut\ and \etacut\ and calorimeter towers with \etacut,  but allow the leading 
jet to be in the region $|\eta({\rm jet}\# 1)|<2$.
}
\end{center}
\label{RDF_HERALHC_fig1}
\end{figure}

\begin{figure}[htbp]
\begin{center}
\includegraphics[scale=0.6]{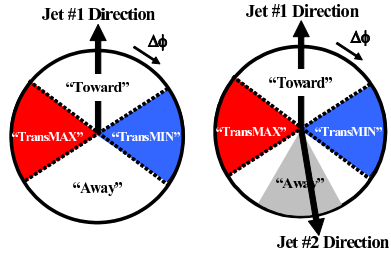}
\caption{
Illustration of correlations in azimuthal angle $\phi$ relative to the direction of the 
leading jet (highest $P_T$ jet) in the event, \Jone.  The angle $\Delta\phi=\phi-\phi_{\rm jet\#1}$ 
is the relative azimuthal angle between charged particles and the direction of \Jone.  On an 
event by event basis, we define \tmax\ (\tmin) to be the maximum (minimum) of the 
two \TR\ regions, $60^\circ<\Delta\phi<120^\circ$ and $60^\circ<-\Delta\phi<120^\circ$.  
\tmax\ and \tmin\ each have an area in \etaphi\ space of $\Delta\eta\Delta\phi=4\pi/6$.  
The overall \TR\ region defined in Fig.~1 contains both the \tmax\ and the \tmin\ regions. Events 
in which there are no restrictions placed on the second 
and third highest $P_T$ jets (\Jtwo\ and \Jthree) are referred to as \LJ\ events ({\it left}).  Events 
with at least two jets with $P_T>15\gevc$ where the leading two jets are nearly \BB\ (\delphicut) with 
$P_T({\rm jet}\#2)/P_T({\rm jet}\#1)>0.8$ and $P_T({\rm jet}\#3)<15\gevc$ are referred to as \BB\ 
events ({\it right}).
}
\end{center}
\label{RDF_HERALHC_fig2}
\end{figure}

\begin{figure}[htbp]
\begin{center}
\includegraphics[scale=0.6]{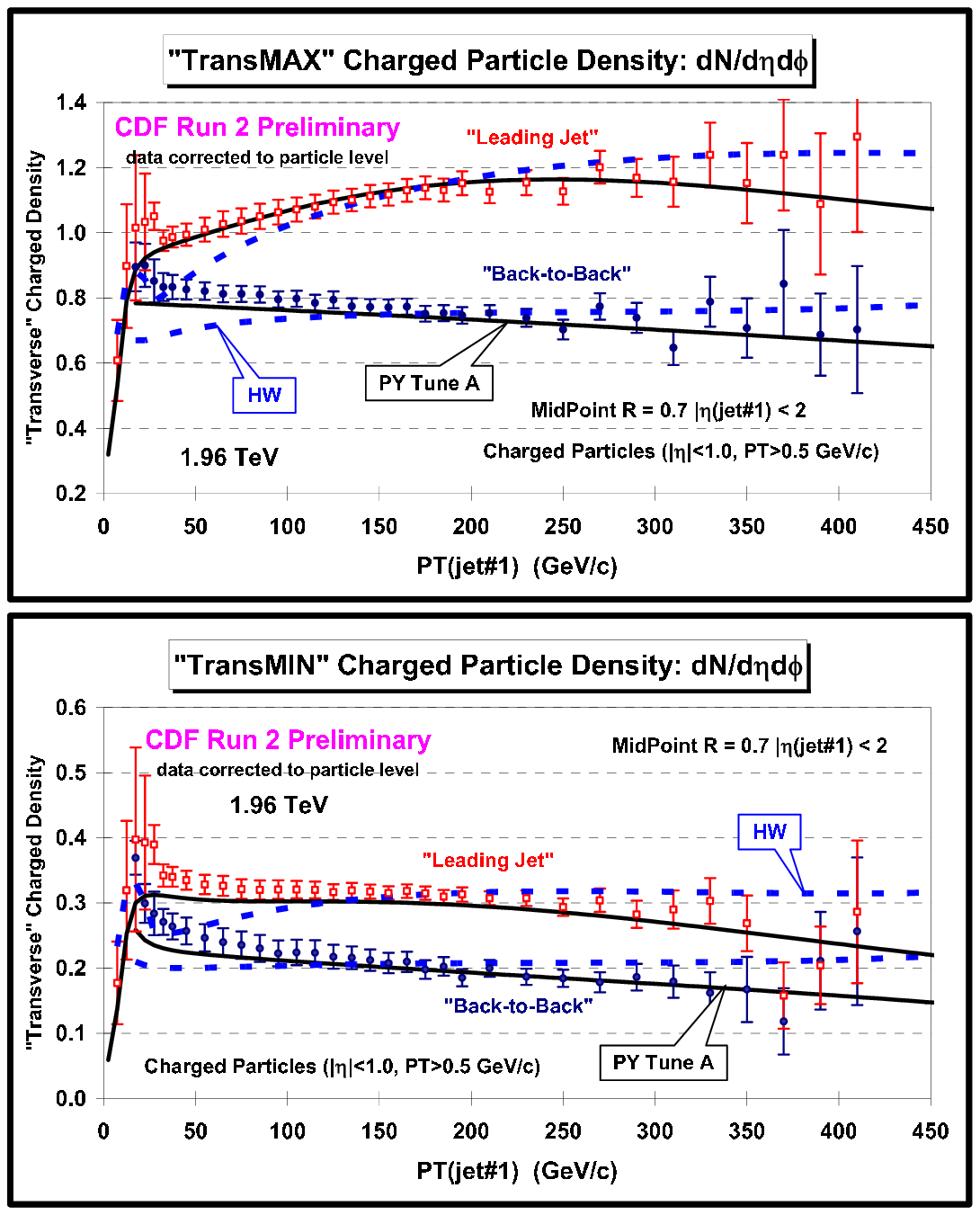}
\caption{
Data at $1.96\tev$ on the density of charged particles, \nden\ with \ptcut\ and \etacut\ in the \tmax\ region ({\it top}) 
and the \tmin\ region ({\it bottom}) for \LJ\ and \BB\ events defined in Fig.~2 as a function of the leading jet 
\bigpt\ compared with PYTHIA Tune A and HERWIG.  The data are corrected to the particle level (with errors that 
include both the statistical error and the systematic uncertainty) and compared with the theory at the 
particle level (\ie generator level).
}
\end{center}
\label{RDF_HERALHC_fig3}
\end{figure}

As illustrated in Fig.~2, we define a variety of MAX and MIN \TR\ regions which helps separate the ``hard component" 
(initial and final-state radiation) from the ``beam-beam remnant" component.  MAX (MIN) refer to the \TR\ region 
containing largest (smallest) number of charged particles or to the region containing the largest (smallest) 
scalar \ptsum\ of charged particles or the region containing the largest (smallest) scalar \etsum\ of particles.  
Since we will be studying regions in \etaphi\ space with different areas, we will construct densities by dividing 
by the area.  For example, the number density, \nden, corresponds the number of charged particles (\ptcut) per 
unit \etaphi\ and the PTsum density, \ptden, corresponds the amount of charged particle (\ptcut) scalar \ptsum\ 
per unit \etaphi, and the transverse energy density, \etden, corresponds the amount of scalar \etsum\ of all 
particles per unit \etaphi. One expects that \tmax\ region will pick up the hardest initial or final-state radiation 
while both the \tmax\ and \tmin\ regions should receive ``beam-beam remnant" contributions.  Hence one expects \tmin\ 
region to be 
more sensitive to the ``beam-beam remnant" component of the \UE, while the \tmax\ minus the \tmin\ (\ie \tdif) 
is very sensitive to hard initial and final-state radiation.  This idea, was first suggested by Bryan Webber, and 
implemented by in a paper by Jon Pumplin \cite{Pumplin}.   Also, Valaria Tano studied this in her CDF Run~1 analysis of 
maximum and minimum transverse cones \cite{CDF:Tano}.

Our previous Run~2 \UE\ analysis \cite{Field:ISMD04} used JetClu to define ``jets" and compared uncorrected data with 
PYTHIA Tune A \cite{PY1,PY2,PY3,PY4,Field:FNAL02} and HERWIG \cite{HW1,HW2} after detector simulation (\ie CDFSIM).  This analysis uses the MidPoint jet 
algorithm ($R=0.7$, $f_{merge}=0.75$) and corrects the observables to the particle level.  The corrected 
observables are then compared with the QCD Monte-Carlo models at the particle level (\ie generator level).   
The models includes PYTHIA Tune A, HERWIG, and a tuned version of JIMMY \cite{JIM}.  In addition, for the first time 
we study the transverse energy density in the \TR\ region.  

\begin{figure}[htbp]
\begin{center}
\includegraphics[scale=0.6]{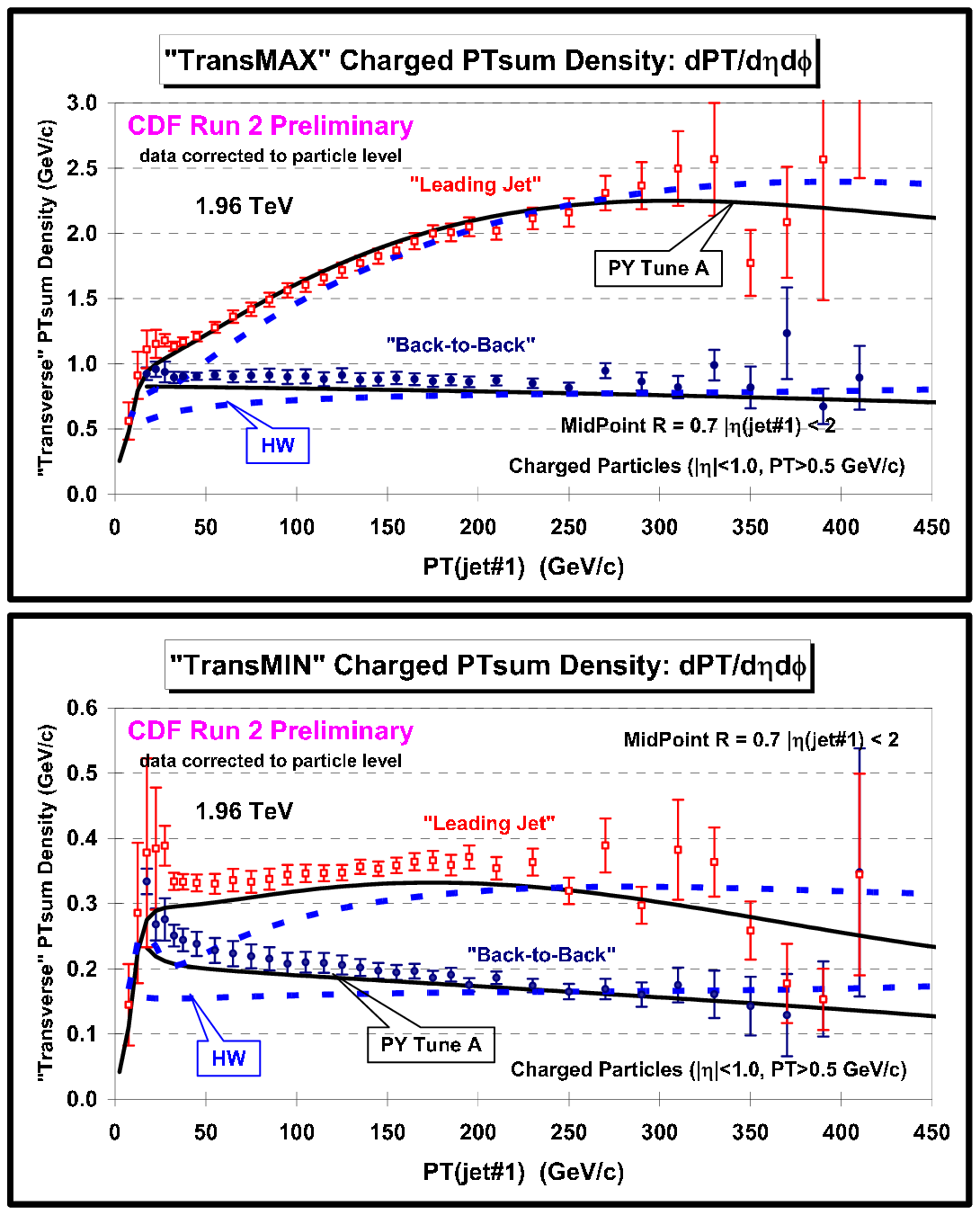}
\caption{
Data at $1.96\tev$ on scalar \ptsum\ density of charged particles, \ptden, with \ptcut\ and \etacut\ in the \tmax\ region ({\it top}) 
and the \tmin\ region ({\it bottom}) for \LJ\ and \BB\ events defined in Fig.~2 as a function of the leading jet 
\bigpt\ compared with PYTHIA Tune A and HERWIG.  The data are corrected to the particle level (with errors that 
include both the statistical error and the systematic uncertainty) and compared with the theory at the 
particle level (\ie generator level).
}
\end{center}
\label{RDF_HERALHC_fig4}
\end{figure}

\begin{figure}[htbp]
\begin{center}
\includegraphics[scale=0.6]{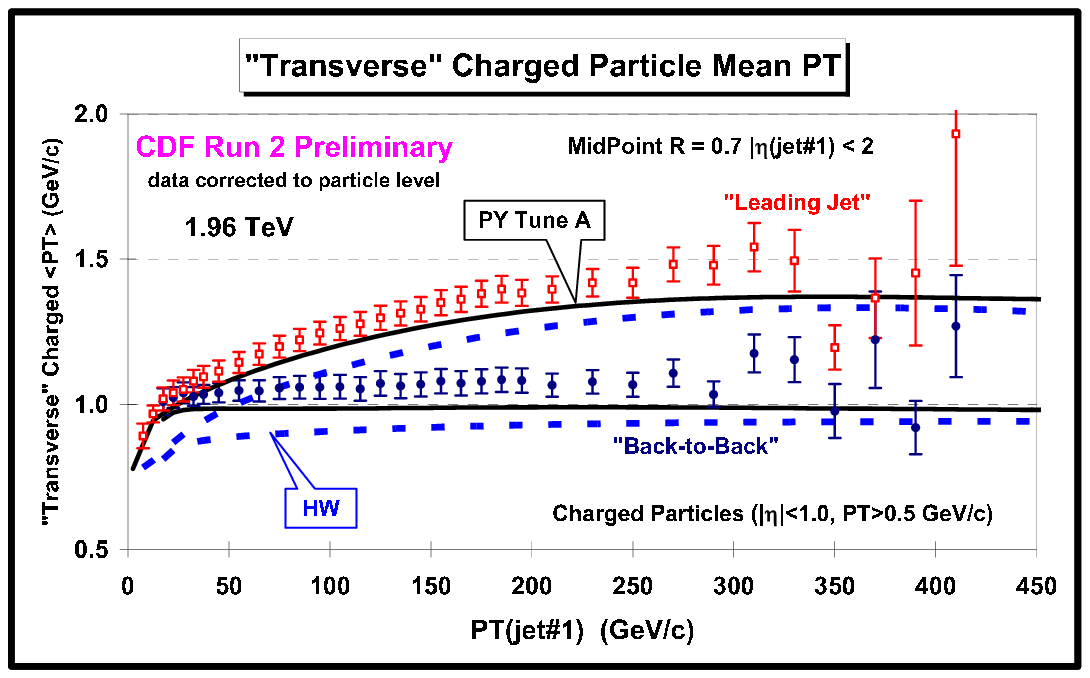}
\caption{
Data at $1.96\tev$ on average transverse momentum, \avept, of charged particles with \ptcut\ and \etacut\ in the 
\TR\ region for \LJ\ and \BB\ events defined in Fig.~2 as a function of the leading jet 
\bigpt\ compared with PYTHIA Tune A and HERWIG.  The data are corrected to the particle level (with errors that 
include both the statistical error and the systematic uncertainty) and compared with the theory at the 
particle level (\ie generator level).
}
\end{center}
\label{RDF_HERALHC_fig5}
\end{figure}

\begin{figure}[htbp]
\begin{center}
\includegraphics[scale=0.6]{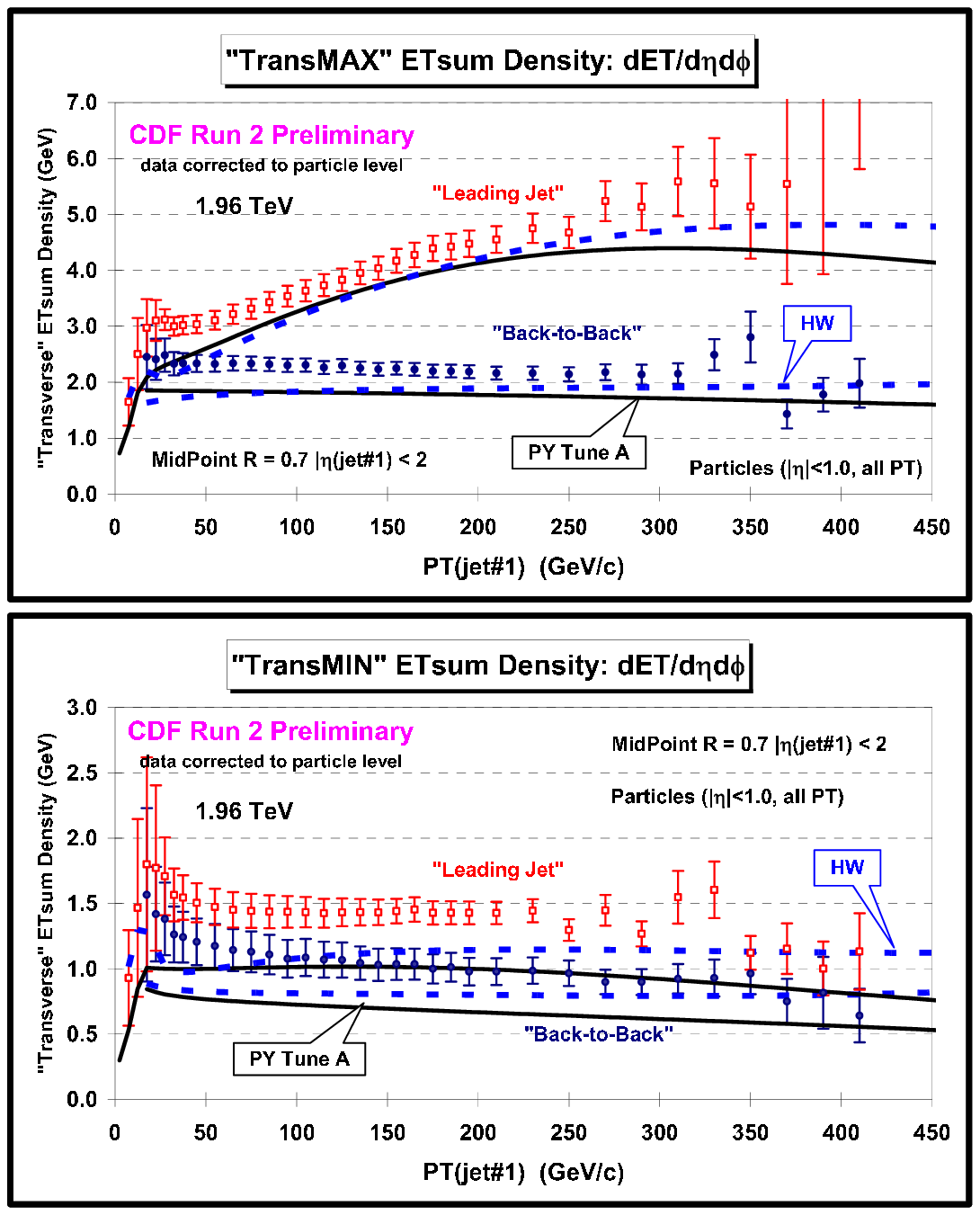}
\caption{
Data at $1.96\tev$ on scalar \etsum\ density, \etden, for particles with \etacut\ in the \tmax\ region ({\it top}) 
and the \tmin\ region ({\it bottom}) for \LJ\ and \BB\ events defined in Fig.~2 as a function of the leading jet 
\bigpt\ compared with PYTHIA Tune A and HERWIG.  The data are corrected to the particle level (with errors that 
include both the statistical error and the systematic uncertainty) and compared with the theory at the 
particle level (\ie generator level).
}
\end{center}
\label{RDF_HERALHC_fig6}
\end{figure}

\begin{figure}[htbp]
\begin{center}
\includegraphics[scale=0.6]{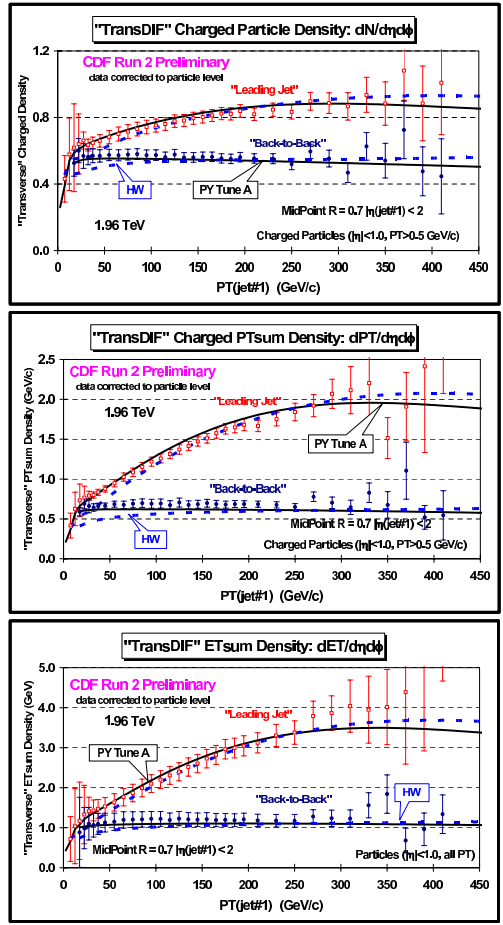}
\caption{
Data at $1.96\tev$ on the difference of the \tmax\ and \tmin\ regions (\tdif\ = \tmax - \tmin)
for \LJ\ and \BB\ events defined in Fig.~2 as a function of the leading jet 
\bigpt\ compared with PYTHIA Tune A and HERWIG.  The data are corrected to the particle level (with errors that 
include both the statistical error and the systematic uncertainty) and compared with the theory at the 
particle level (\ie generator level).
}
\end{center}
\label{RDF_HERALHC_fig7}
\end{figure}

\begin{figure}[htbp]
\begin{center}
\includegraphics[scale=0.6]{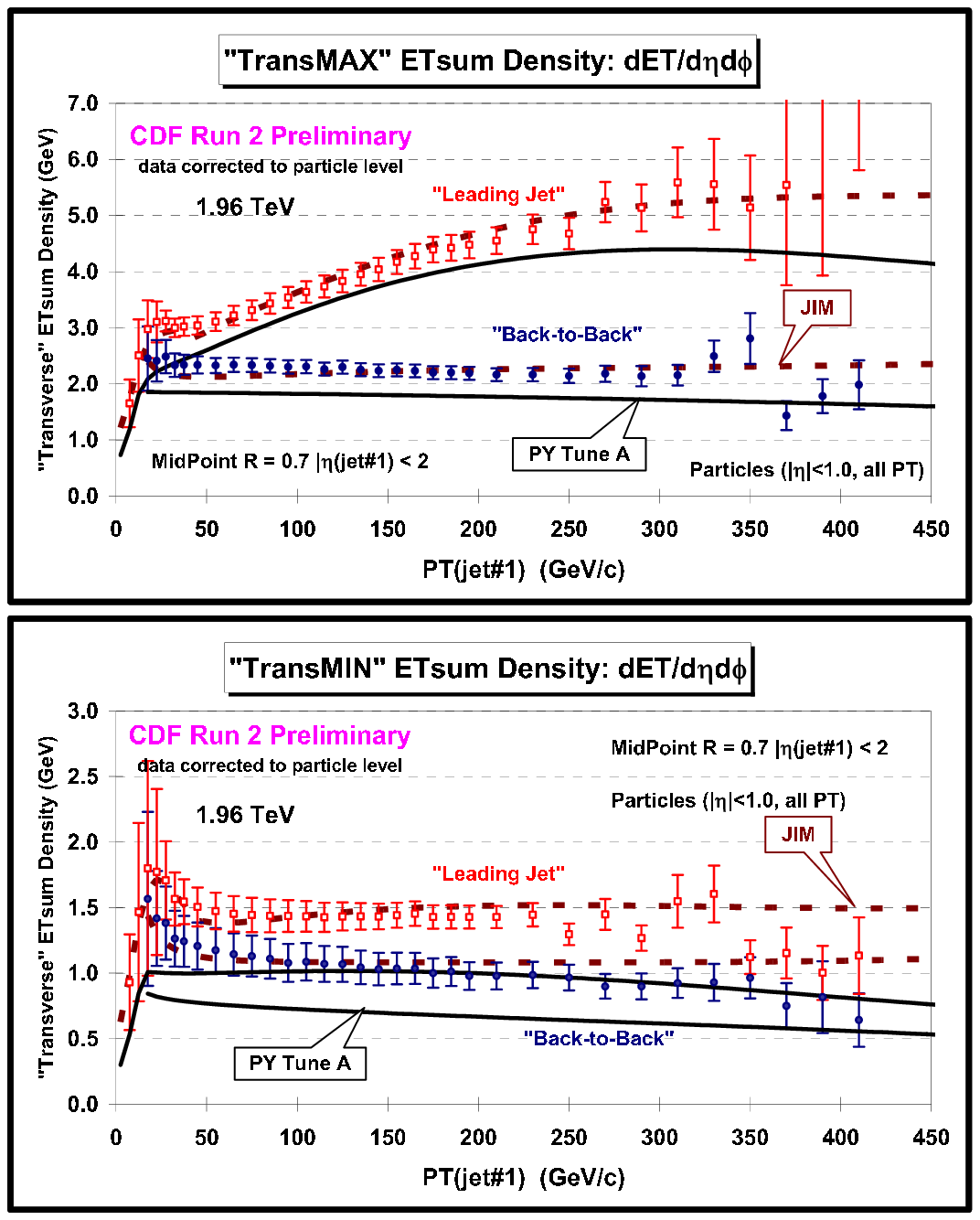}
\caption{
Data at $1.96\tev$ on scalar \etsum\ density, \etden, for particles with \etacut\ in the \tmax\ region ({\it top}) 
and the \tmin\ region ({\it bottom}) for \LJ\ and \BB\ events defined in Fig.~2 as a function of the leading jet 
\bigpt\ compared with PYTHIA Tune A and tuned JIMMY. JIMMY was tuned to fit the \TR\ energy density in \LJ\ events 
($PTJIM=3.25\gevc$). The data are corrected to the particle level (with errors that 
include both the statistical error and the systematic uncertainty) and compared with the theory at the 
particle level (\ie generator level).
}
\end{center}
\label{RDF_HERALHC_fig8}
\end{figure}

\begin{figure}[htbp]
\begin{center}
\includegraphics[scale=0.6]{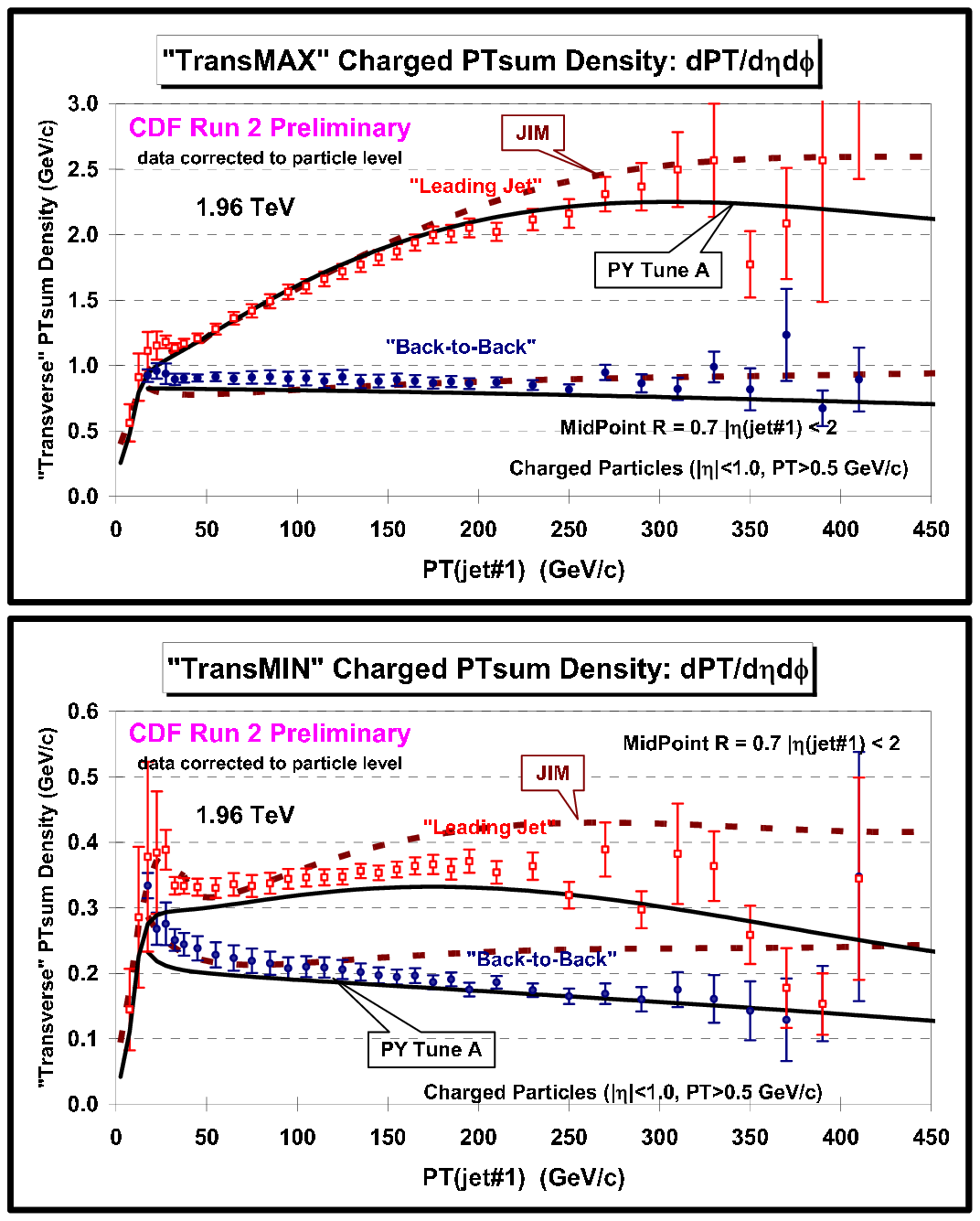}
\caption{
Data at $1.96\tev$ on scalar \ptsum\ density of charged particles, \ptden, with \ptcut\ and \etacut\ in the \tmax\ region ({\it top}) 
and the \tmin\ region ({\it bottom}) for \LJ\ and \BB\ events defined in Fig.~2 as a function of the leading jet 
\bigpt\ compared with PYTHIA Tune A and tuned JIMMY. JIMMY was tuned to fit the \TR\ energy density in \LJ\ events 
($PTJIM=3.25\gevc$). The data are corrected to the particle level (with errors that 
include both the statistical error and the systematic uncertainty) and compared with the theory at the 
particle level (\ie generator level).
}
\end{center}
\label{RDF_HERALHC_fig9}
\end{figure}

\begin{figure}[htbp]
\begin{center}
\includegraphics[scale=0.6]{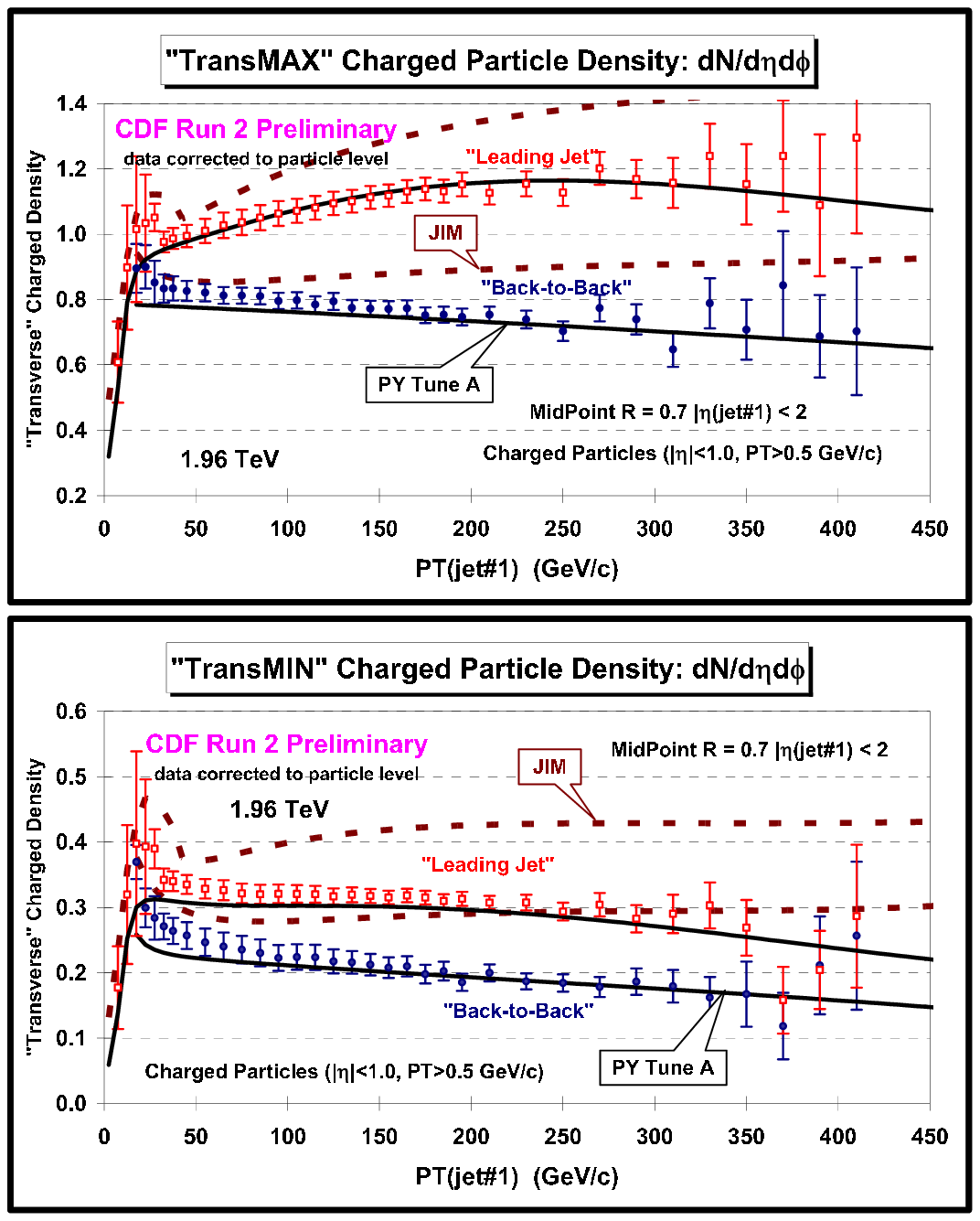}
\caption{
Data at $1.96\tev$ on the density of charged particles, \nden, with \ptcut\ and \etacut\ in the \tmax\ region ({\it top}) 
and the \tmin\ region ({\it bottom}) for \LJ\ and \BB\ events defined in Fig.~2 as a function of the leading jet 
\bigpt\ compared with PYTHIA Tune A and tuned JIMMY. JIMMY was tuned to fit the \TR\ energy density in \LJ\ events 
($PTJIM=3.25\gevc$). The data are corrected to the particle level (with errors that 
include both the statistical error and the systematic uncertainty) and compared with the theory at the 
particle level (\ie generator level).
}
\end{center}
\label{RDF_HERALHC_fig10}
\end{figure}

\begin{figure}[htbp]
\begin{center}
\includegraphics[scale=0.6]{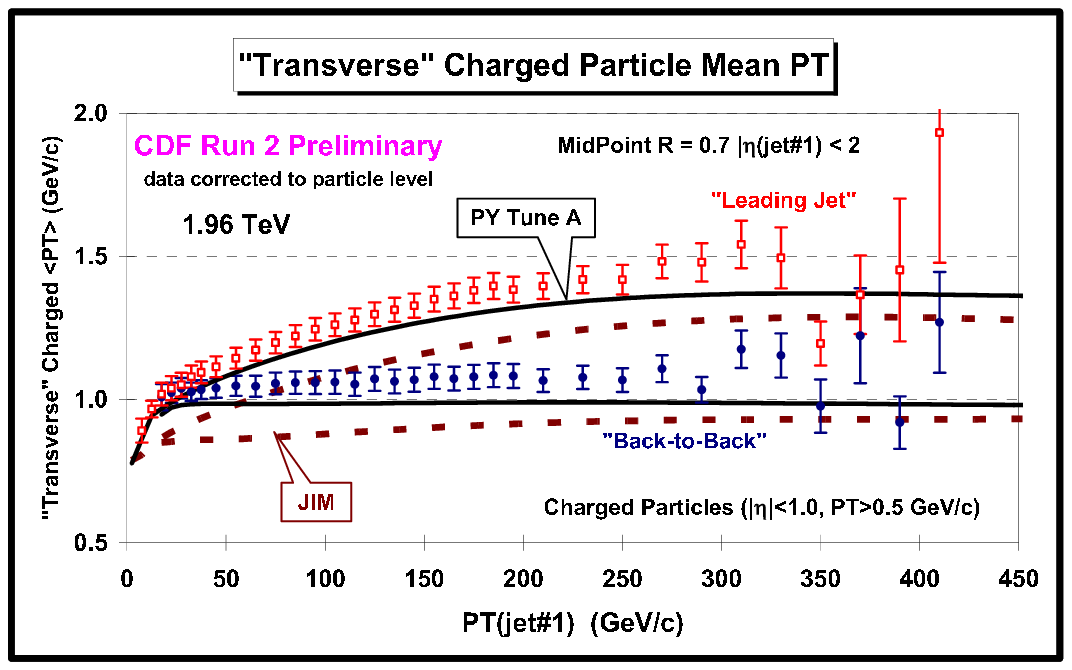}
\caption{
Data at $1.96\tev$ on average transverse momentum, \avept, of charged particles with \ptcut\ and \etacut\ in the 
\TR\ region for \LJ\ and \BB\ events defined in Fig.~2 as a function of the leading jet 
\bigpt\ compared with PYTHIA Tune A and tuned JIMMY. JIMMY was tuned to fit the \TR\ energy density in \LJ\ events 
($PTJIM=3.25\gevc$). The data are corrected to the particle level (with errors that 
include both the statistical error and the systematic uncertainty) and compared with the theory at the 
particle level (\ie generator level).
}
\end{center}
\label{RDF_HERALHC_fig11}
\end{figure}

Fig.~3 and Fig.~4 compare the data on the density of charged particles and the charged \ptsum\ density in the 
\TR\ region corrected to the particle level for \LJ\ and \BB\ events with PYTHIA Tune A and HERWIG at the 
particle level. As expected, the \LJ\ and \BB\ events behave quite differently.  For the \LJ\ case the 
\tmax\ densities rise with increasing \ptone, while for the \BB\ case they fall with increasing \ptone.  
The rise in the \LJ\ case is, of course, due to hard initial and final-state radiation, which has been suppressed 
in the \BB\ events.  The \BB\ events allow for a more close look at the ``beam-beam remnant" and 
multiple parton scattering component of the \UE\ and PYTHIA Tune A (with multiple parton interactions) does a 
better job describing the data than HERWIG (without multiple parton interactions).  

The \tmin\ densities are more sensitive to the ``beam-beam remnant"  and multiple parton interaction component 
of the \UE.  The \BB\ data show a decrease in the \tmin\ densities with increasing \ptone\ which is described 
fairly well by PYTHIA Tune A (with multiple parton interactions) but not by HERWIG (without multiple parton 
interactions).  The decrease of the \tmin\ densities with increasing \ptone\ for the \BB\ events is very 
interesting and might be due to a ``saturation" of the multiple parton interactions at small impact parameter.  
Such an effect is included in PYTHIA Tune A but not in HERWIG (without multiple parton interactions).

Fig.~5 compares the data on average \pt\ of charged particles in the \TR\ region corrected to the particle level 
for \LJ\ and \BB\ events with PYTHIA Tune A and HERWIG at the particle level.    Again the \LJ\ and \BB\ events 
behave quite differently. 
 
Fig.~6 shows the data corrected to the particle level for the scalar \etsum\ density in the \TR\ region for 
\LJ\ and \BB\ events compared with PYTHIA Tune A and HERWIG.  The scalar \etsum\ density 
has been corrected to correcpond to all particles (all \pt, \etacut). Neither PYTHIA Tune A nor HERWIG produce 
enough energy in the \TR\ region.  HERWIG has more ``soft" particles than PYTHIA Tune A and does slightly better 
in describing the energy density in the \tmax\ and \tmin\ regions.

Fig.~7 shows the difference of the \tmax\ and \tmin\ regions (\tdif\ $=$ \tmax\ minus \tmin) for 
\LJ\ and \BB\ events compared with PYTHIA Tune A and HERWIG.  ``TransDIF" is more sensitive to the hard 
scattering component of the \UE\ (\ie initial and final state radiation).  Both PYTHIA Tune A and HERWIG 
underestimate the energy density in the \tmax\ and \tmin\ regions (see Fig.~6).  However, they both fit 
the \tdif\ energy density.  This indicates that the excess energy density seen in the data probably 
arises from the ``soft" component of the \UE\ (\ie beam-beam remnants and/or multiple parton interactions). 

JIMMY is a model of multiple parton interaction which can be combined with HERWIG to enhance the 
\UE\ thereby improving the agreement with data. Fig.~8 and Fig.~9 shows the energy density and charged \ptsum\ 
density, respectively, in the \tmax\ and \tmin\ regions for \LJ\ and \BB\ events compared with PYTHIA Tune A and 
a tuned version of JIMMY.  JIMMY was tuned to fit the \TR\ energy density in \LJ\ events ($PTJIM=3.25\gevc$). 
The default JIMMY ($PTJIM=2.5\gevc$) produces too much energy and too much charged \ptsum\ in the \TR\ region. 
Tuned JIMMY does a good job of fitting the energy and charged \ptsum\ density in the \TR\ region (although it 
produces slightly too much charged PTsum at large \ptone).  However, the tuned JIMMY produces too many 
charged particles with \ptcut\ (see Fig. 10).  The particles produced by this tune of JIMMY are too soft.  
This can be seen clearly in Fig.~11 which shows the average charge particle \pt\ in the \TR\ region.

The goal of this analysis is to produce data on the \UE\ that is corrected to the particle level so that it 
can be used to tune the QCD Monte-Carlo models without requiring CDF detector simulation.  Comparing the 
corrected observables with PYTHIA Tune A and HERWIG at the particle level (\ie generator level) leads to the 
same conclusions as we found when comparing the uncorrected data with the Monte-Carlo models after detector 
simulation \cite{Field:ISMD04}.  PYTHIA Tune A (with multiple parton interactions) does a better job in describing the 
\UE\ (\ie \TR\ regions) for both \LJ\ and \BB\ events than does HERWIG (without multiple parton interactions).   
HERWIG does not have enough activity in the \UE\ for \ptone\ less than about $150\gevc$, which was also 
observed in our published Run~1 analysis \cite{CDF:Field}.

This analysis gives our first look at the energy in the \UE\ (\ie the \TR\ region).  Neither PYTHIA Tune A nor 
HERWIG produce enough transverse energy in the \TR\ region.  However, they both fit the \tdif\ energy 
density (\tmax\ minus \tmin).  This indicates that the excess energy density seen in the data probably 
arises from the ``soft" component of the \UE\ (\ie beam-beam remnants and/or multiple parton interactions). 
HERWIG has more ``soft" particles than PYTHIA Tune A and does slightly better in describing the energy density 
in the \tmax\ and \tmin\ regions.  Tuned JIMMY does a good job of fitting the energy and charged \ptsum\ density 
in the \TR\ region (although it produces slightly too much charged \ptsum\ at large \ptone).  However, the 
tuned JIMMY produces too many charged particles with \ptcut\ indicating that the particles produced by 
this tuned JIMMY are too soft.  

In summary, we see interesting dependence of the \UE\ on the transverse momentum of the leading jet 
(\ie the $Q^2$ of the hard scattering).  For the \LJ\ case the \tmax\ densities rise with increasing \ptone, while 
for the \BB\ case they fall with increasing \ptone.  The rise in the \LJ\ case is due to hard initial and final-state 
radiation with $p_T> 15\gevc$, which has been suppressed in the \BB\ events.  The \BB\ data show a decrease in the \tmin\ 
densities with increasing \ptone. The decrease of the \tmin\ densities with increasing \ptone\ for the \BB\ events is 
very interesting and might be due to a ``saturation" of the multiple parton interactions at small impact parameter.  
Such an effect is included in PYTHIA Tune A (with multiple parton interactions) but not in HERWIG (without multiple 
parton interactions).  PYTHIA Tune A does predict this decrease, while HERWIG shows an increase (due to increasing initial 
and final state radiation).


\end{document}